\documentclass{taira}

\usepackage{graphicx}
\usepackage[percent]{overpic}
\usepackage{epstopdf, epsfig}
\usepackage{graphicx}
\usepackage{dcolumn}
\usepackage{bm}
\usepackage{color}
\usepackage{array}
\usepackage{amsmath,amssymb,stmaryrd} 
\usepackage{soul} 
\usepackage{xcolor}
\usepackage{lscape}
\usepackage{booktabs} 
\usepackage{xcolor}
\usepackage{tikz}
\usetikzlibrary{shapes}

\definecolor{blue3}{rgb}{0, 0.1770, 0.3410}

\newcommand{\RedCircle}{\raisebox{0.0pt}{\tikz{\node[scale=0.7,regular polygon, circle, draw = {rgb,255:red,255; green,0; blue,0},line width=0.3mm, fill={rgb,255:red,255; green,0; blue,0}](){};}}}
\newcommand{\RedCircleEmpty}{\raisebox{0.0pt}{\tikz{\node[scale=0.7,regular polygon, circle, draw = {rgb,255:red,255; green,0; blue,0},line width=0.15mm, fill={rgb,255:red,255; green,255; blue,255}](){};}}}
\newcommand{\GrayDownTriangle}{\raisebox{0.0pt}{\tikz{\node[scale=0.45,regular polygon, regular polygon sides=3,fill={rgb,255:red,64; green,64; blue,64},rotate=-180](){};}}}
\newcommand{\BlueSqaure}{\raisebox{0.0pt}{\tikz{\node[scale=0.62,regular polygon, regular polygon sides=4, draw = {rgb,255:red,0; green,0; blue,255}, line width=0.3mm, fill={rgb,255:red,0; green,0; blue,255},rotate=0](){};}}}

\begin{document}  

\newtheorem{lemma}{Lemma}
\newtheorem{corollary}{Corollary}

\shorttitle{Wake interactions between two side-by-side cylinders} 
\shortauthor{K. Zhang \& M. N. Haque } 

\title{Wake interactions between two side-by-side circular cylinders with different sizes}

\author
 {
 Kai Zhang\aff{1,2} \corresp{\email{kai.zhang3@rutgers.edu}}
\and
 Md. Naimul Haque\aff{3}
}

\affiliation
{
\aff{1} School of Naval Architecture, Ocean and Civil Engineering, Shanghai Jiao Tong University, Shanghai 200240, China
\aff{2} Department of Mechanical and Aerospace Engineering, Rutgers University, Piscataway, NJ 08854, USA 
\aff{3} Department of Civil Engineering, East West University, Dhaka-1212, Bangladesh
}

\maketitle
\begin{abstract}
Flows over two side-by-side circular cylinders exhibit fascinating flow physics due to complex interactions between the coupled wakes.
However, their mutual interference effects have not been elucidated in a quantitative manner thus far.
In this paper, we create a mismatch between the two wakes by introducing a size difference in the cylinder pair, such that the effects of one wake on the other can be distinguished.
Depending on the size and gap ratios between the two cylinders, the coupled wake exhibits distinct dynamical features including the quasi-periodic flow, synchronized flow, and chaotic flow.
Through advanced spectral analysis of lift coefficients and dynamic mode decomposition of the flow fields, we reveal that the quasi-periodic flows are mainly composed of two primary frequencies associated with vortex shedding in the near wakes of the two cylinders.
Both wakes impose their own frequencies on the other, resulting in the beating phenomenon in the lift coefficients.
The triad interactions between the two wakes generate the sideband frequencies, which are associated with modal structures that are mostly active in the far wake.
The transition from quasi-periodicity to synchronization is dominated by the vortex shedding behind the larger cylinder, to which the wake of the smaller cylinder locks in.
These results reveal new insights on the coupled wakes of two circular cylinders, and are pivotal for understanding more general wake interaction problems.

\end{abstract}
\section{Introduction}

Flow over multiple bluff bodies in proximity is a common occurrence in various engineering applications such as high-rise buildings, marine structures, heat exchange tubes, just to name a few. 
The interactions between multiple wakes give rise to complex flow patterns, which are often associated with substantial flow-induced forces, sound generation, and structural vibrations \citep{zdravkovich1997flow,sumner2010two,zhou2016wake}. 
As a prototype of this general flow interference problem, the simplified configuration of flow over two side-by-side cylinders has been investigated extensively.
Most of these studies have focused on the case where the two cylinders are the same in size.
 
The wake dynamics of a pair of side-by-side identical cylinders is significantly influenced by the gap distance between the two bodies \citep{williamson1985evolution,peschard1996coupled,kang2003characteristics}. 
For cylinders separated with large distance (typically larger than 5 cylinder diameters), the mutual interference between the two wakes is negligible, leading to two separate single-cylinder wakes.
As the gap distance is reduced, the two vortex streets become synchronized. 
With relatively large gap, the wake is characterized by two parallel vortex streets with anti-phase shedding \citep{bearman1973interaction,williamson1985evolution}. 
With smaller gap, the wake transitions to in-phase shedding, and the two vortex streets merge into a ``binary vortex" at downstream \citep{williamson1985evolution}.
The flow between the two cylinders becomes more prominent as the gap further decreases.
The wake becomes asymmetric, and is characterized by a deflected gap flow with the formation of a narrow and a wide wake. 
The gap flow is observed to switch randomly between the two opposite directions, leading to the ``flip-flop'' phenomenon \citep{kim1988investigation,carini2014origin}.
As the two cylinders are brought into contact, a single vortex street forms in the wake, with the characteristic length scale being double the diameter of the cylinder.
The transitions between the above different flow states are also found to be associated with bistabilities \citep{kang2003characteristics,mizushima2008stability,ren2021bistabilities}.

Despite the vast literature on the flows over two identical cylinders, a deep understanding of how one wake is affected by the other is hindered by the difficulty in differentiating the neighbouring wake's interference effects from its own wake dynamics, since both have the same frequency.
In this study, we break such symmetry by introducing a size difference in the cylinder pair.
This creates a scenario where the two wakes with different shedding frequencies and strengths are coupled, such that their mutual interaction effects can be distinguished through spectral analysis.
Among the few who have studied a similar problem setup, \citet{inoue2008beat} numerically investigated the sound generation by two square cylinders of different sizes placed in a side-by-side arrangement at a low Reynolds number of 150. 
\cite{octavianty2015synchronized} further studied the vortex shedding from two side-by-side rectangular cylinders with different cross-sectional aspect ratios of 1.2--1.5 at Reynolds number of the order $10^4$.
Both works revealed interesting flow phenomenon such as beating and lock-in.
However, the discussions in these studies remain qualitative, and a thorough understanding of the wake interference effects is not achieved.

Aiming at providing an improved understanding of this wake interference problem, we carry out a large number of direct numerical simulations of flows over two side-by-side cylinders with difference sizes and gap distances. 
We quantify the mutual interactions between the parallel wakes by performing spectral analysis and dynamic mode decomposition (DMD) to the flows, and reveal novel flow physics that can not be observed in the wakes of two identical cylinders.
The insights obtained from this study should improve the understanding of more general wake interaction problems.
In what follows, we provide details of computational setup in \S \ref{sec:setup}. 
The results are presented in \S \ref{sec:results}.
We conclude this study by summarizing our findings in \S \ref{sec:conclusion}.

\section{Computational setup}
\label{sec:setup}
\begin{figure}
    \centering
    \includegraphics[scale=0.55]{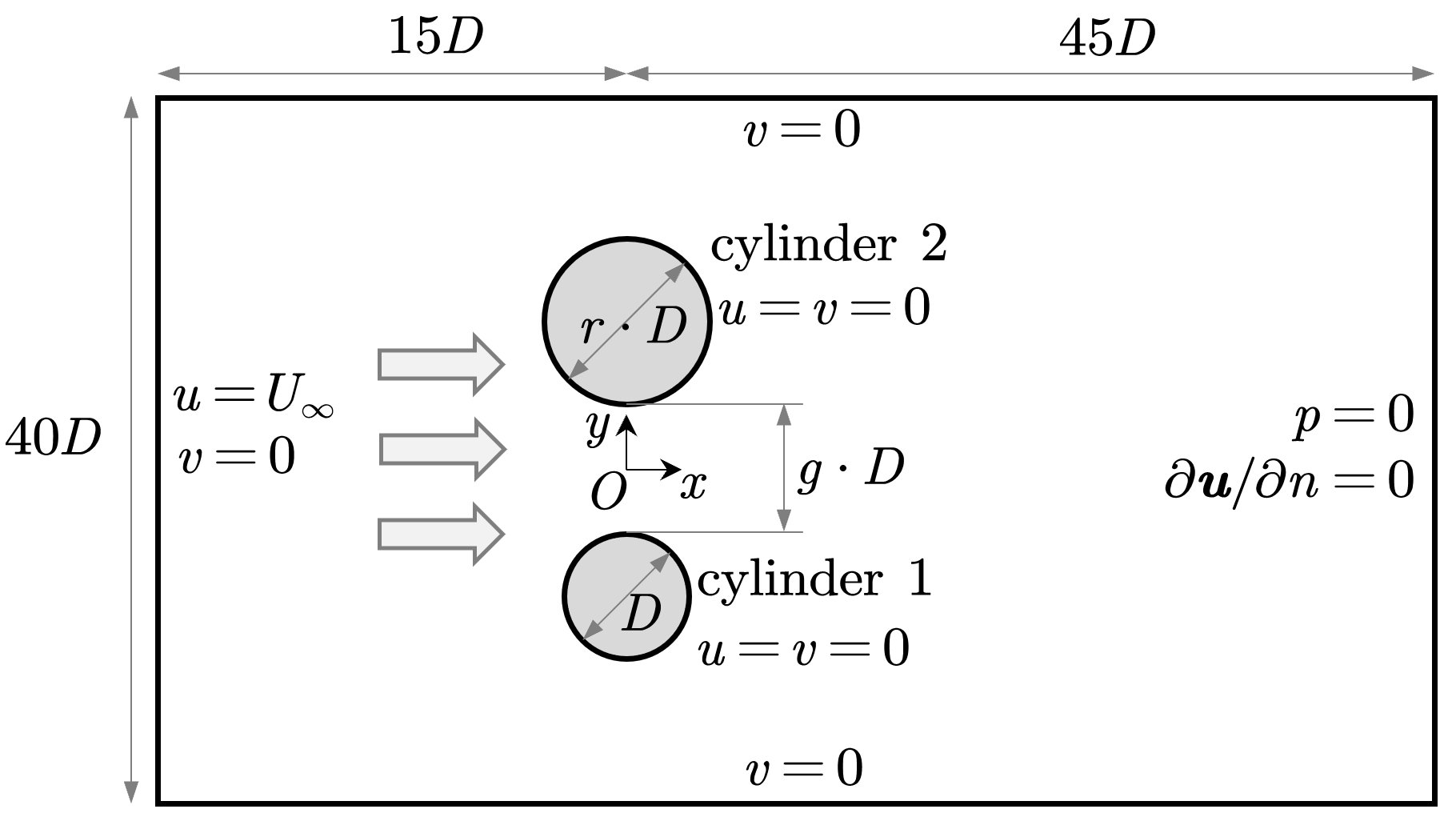}
    \caption{Schematic of the computational setup. The sketch is not to scale.}
    \label{fig:schematic}
\end{figure}

The problem of interest is schematically shown in figure \ref{fig:schematic}. 
Two cylinders with diameters $D$ and $rD$, where $r$ is the nondimensional size ratio, are placed side-by-side to the uniform inflow velocity $U_{\infty}$. 
The gap distance between the two cylinders is denoted as $gD$, with $g$ being the nondimensional gap ratio. 
In this study, we consider a range of size ratio $r\in[1/1.3, 1.3]$, and gap ratio $g\in[1,3]$.
The Reynolds number is defined based on cylinder 1, i.e., $Re_D\equiv U_{\infty}D/\nu$, and is fixed at 100. 
In what follows, the spatial variables are normalized by $D$, velocity by $U_{\infty}$, and time by $D/U_{\infty}$.

A finite-volume-based incompressible flow solver \emph{pimpleFoam} from the OpenFOAM package \citep{weller1998tensorial} is used for solving the two-dimensional Navier-Stokes equations with second-order-accuracies in both space and time.
The rectangular computational domain covers $(x,y)\in [-15,45]\times[-20,20]$.
The wake region behind the cylinder pair is discretized with uniform grid.
The inlet boundary is prescribed with the freestream velocity $(u,v)=(U_{\infty},0)$, where $u$ and $v$ are the components in the velocity vector $\boldsymbol{u}$.
The surfaces of the cylinders are treated as no-slip $(u,v)=(0,0)$.
The side boundaries are considered as slip $v=0$.
The zero-gradient condition is applied to the outlet for velocity, where a reference pressure $p=0$ is specified.
The drag, lift coefficients and the Strouhal number are defined based on the diameter of cylinder 1 as 
\begin{equation}
C_d = \frac{F_d}{\rho U_{\infty}^2 D/2 }, \quad C_l = \frac{F_l}{\rho U_{\infty}^2 D/2 }, \quad \textrm{and} \quad St = \frac{fD}{U_{\infty}},
\label{equ:CdClSt}
\end{equation}
where $F_d$ and $F_l$ are the drag and lift force on the cylinder with unit length, $\rho$ is the fluid density, and $f$ is the frequency.
These definitions allow a straightforward comparison of the hydrodynamic quantities between two cylinders.

\begin{table}
\renewcommand*{\arraystretch}{1.2}
 \begin{center}
  \begin{tabular}{l@{\hskip 0.7cm}|@{\hskip 0.7cm}c@{\hskip 0.7cm}c@{\hskip 0.7cm}c@{\hskip 0.7cm}c}
    		& $\overline{C_d}$  &   $\overline{C_l} $ & $C_l^{\prime}$  &  $St $  \\
    		\midrule
				\citet{kang2003characteristics}   & 1.434 & 0.178 & 0.192 & 0.164\\
    		\citet{lee2009flow} 								& 1.452 & 0.185 & 0.198 &	---	 \\
				\citet{carini2014origin}   				& 1.409 & --- 		& 0.185 & 0.163 \\

				Present 															& 1.471 & 0.176 & 0.195 & 0.165 \\
				Refined mesh 												& 1.469 & 0.175 & 0.194 & 0.165 
  \end{tabular}
  \caption{Comparison of mean drag coefficient ($\overline{C_d}$), mean lift coefficient ($\overline{C_l}$), root-mean-squared lift coefficient ($C_l^{\prime}$) and Strouhal number ($St$) for the in-phase synchronized vortex shedding at $(r,g)=(1,1.5)$. In the refined mesh, the grid resolutions are doubled in both $x$ and $y$ directions.}
 \end{center}
 \label{table:validation}
\end{table}

\begin{figure}
\centering
\includegraphics[width=1\textwidth]{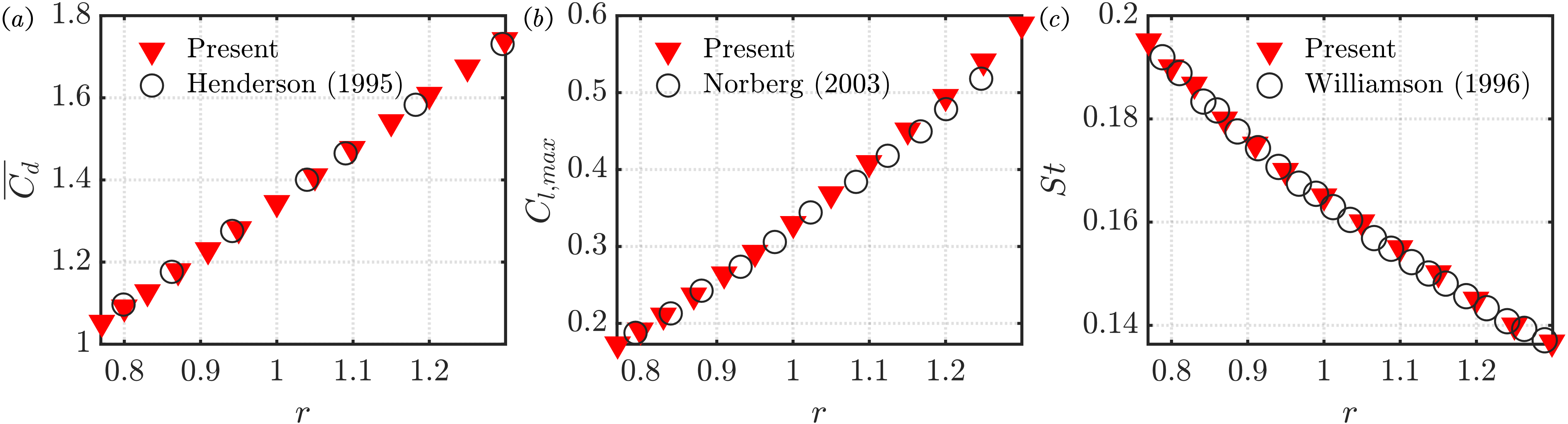}
\caption{Comparison of forces coefficients and Strouhal number of a single cylinder as a function of size ratio $r$. ($a$) mean drag coefficient $\overline{C_d}$ , $(b)$ maximum lift coefficient $C_{l,max}$ and ($c$) Strouhal number $St$. Note that $C_d$, $C_l$ and $St$ are defined based on $D$ (corresponding to $Re=100$ in this study), as stated in equation (\ref{equ:CdClSt}).}
\label{fig:singleCylinder}
\end{figure}

To validate the computational setup, we compare the drag, lift coefficients and the Strouhal number of the case $(r,g)=(1,1.5)$ with previous studies. As shown in table \ref{table:validation}, the key hydrodynamic quantities are in good agreement with the data from the literature.
It is also seen that the refined mesh does not significantly affect the results.
In addition, we compare the key hydrodynamic quantities of flow over a single cylinder with \citet{henderson1995details}, \citet{norberg2003fluctuating} and \citet{williamson1996vortex} in figure \ref{fig:singleCylinder}. 
Again, the results achieve almost perfect match, demonstrating high accuracy of the computational setup.

\section{Results}
\label{sec:results}
As observed from figure \ref{fig:singleCylinder}, with increasing diameter of the cylinder, the natural vortex shedding frequency decreases, and the strength of the vortex shedding (indicated by $C_{l,max}$) increases.
In this section, we discuss the behaviors of the coupled wakes with distinct natural frequencies and strengths.
We first present a classification of flows based on their wake dynamics over a range of size and gap ratios.
The mutual interference effects between the two wakes are then quantified through an advanced spectral analysis of the lift coefficients.
At last, we reveal the spatial modes of interaction in the coupled wakes using the technique of dynamic mode decomposition.

\subsection{Classification of the coupled wakes}
\label{sec:classification}

\begin{figure}
    \centering
    \includegraphics[width=0.95\textwidth]{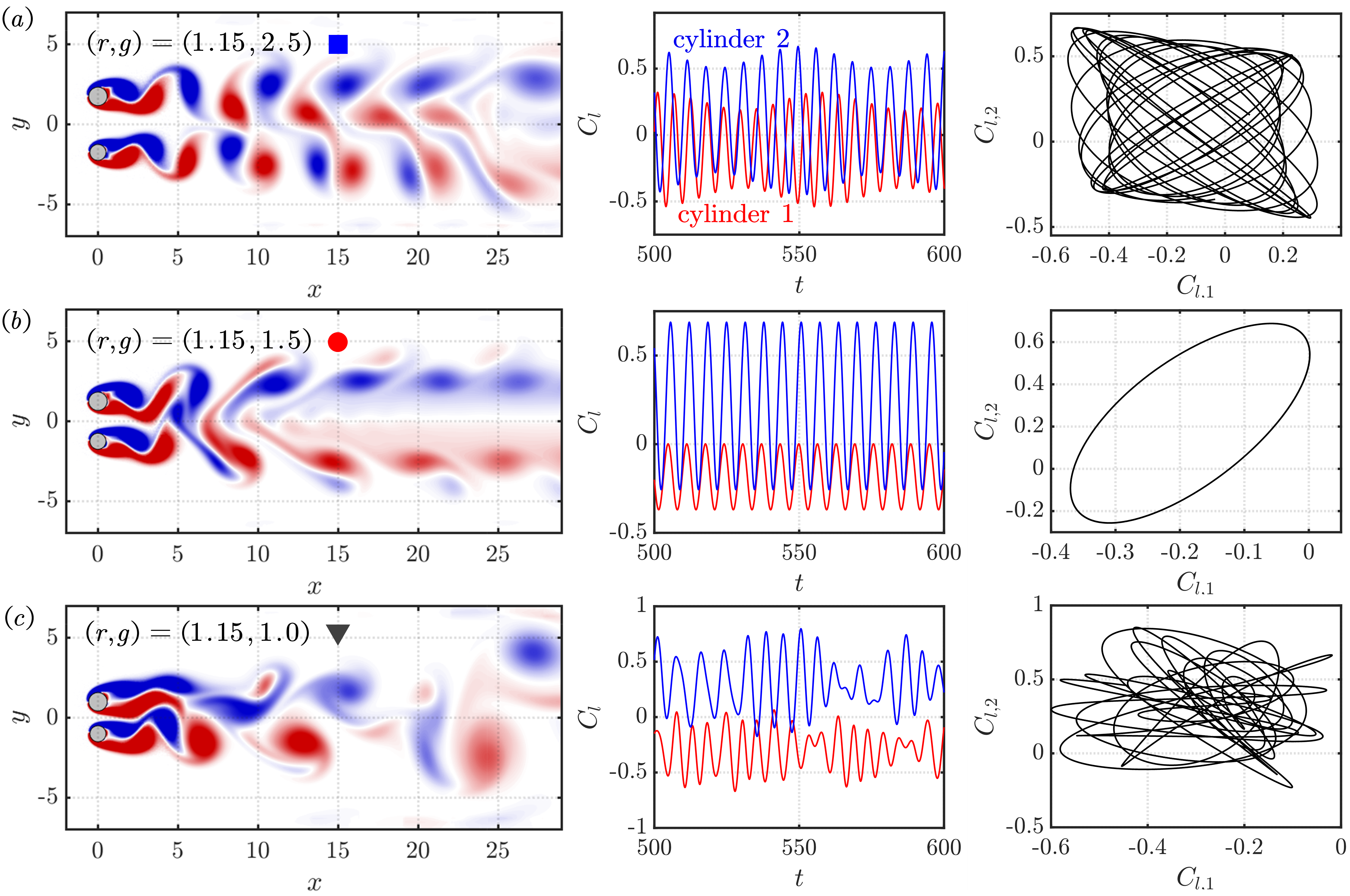}
    \caption{Representative wakes for flow over two cylinders of different sizes. ($a$) quasi-periodic wake at $(r,g)=(1.15,2.5)$, ($b$) synchronized wake at $(r,g)=(1.15,1.5)$ and ($c$) chaotic wake at $(r,g)=(1.15,1.0)$. The left column shows the wake vortical structures visualized by vorticity ($\omega_z=\boldsymbol{\nabla}\times\boldsymbol{u}$) contours ranging from -1 (blue) to 1 (red). The middle column shows the time histories of the lift coefficients, and the right column shows their phase trajectories.}
\label{fig:allFlow}
\end{figure}

\begin{figure}
    \centering
    \includegraphics[width=0.45\textwidth]{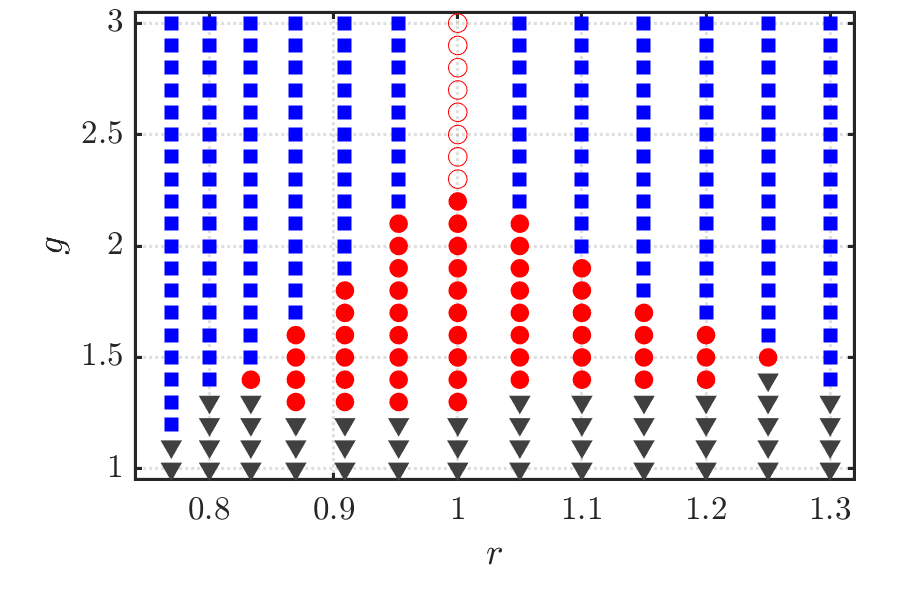}
    \caption{Classification of the flow patterns in the $r$-$g$ space. \protect\BlueSqaure: quasi-periodic wakes; \protect\RedCircle: synchronized wakes; \protect\RedCircleEmpty: anti-phase synchronized vortex shedding (for $r=1$ only);  \protect\GrayDownTriangle: chaotic wakes. }
\label{fig:regimeMap}
\end{figure}

Depending on the size and gap ratios, the flows over two side-by-side uneven cylinders exhibit different dynamical features, including quasi-periodicity, synchronization, and chaotic shedding.
We present these three typical wakes in figure \ref{fig:allFlow}, with their distributions in the $r$-$g$ space mapped out in figure \ref{fig:regimeMap}.
The quasi-periodic flows are mostly observed for cases with large gap ratios, where the mutual interactions between the two wakes are relatively weak.
These flows feature two rows of vortex streets developing separately in the near wakes, as shown in figure \ref{fig:allFlow}($a$).
The shed vortices undergo complex interactions as they are convected further downstream.
The lift coefficients of the two cylinders exhibit beating phenomenon, suggesting the existence of multiple frequency components in the flows.
This also results in a recurrent but not repetitive trajectory in the phase portrait spanned by $C_{l,1}$ and $C_{l,2}$.

Decreasing the gap ratio strengthens the mutual interactions between the two wakes, leading to synchronization of the parallel vortex streets as shown in figure \ref{fig:allFlow}($b$).
The synchronized flow features a C-shaped wake, where the like-sign vortices from the two wakes merge into one larger vortex streets.
The lift coefficients of the two cylinders exhibit periodic oscillations with a closed loop in their phase portrait.
Different from the quasi-periodic flows where the lift fluctuations of the two cylinders are comparable, in the synchronized flow the lift fluctuation on cylinder 2 is significantly larger than that on cylinder 1.
Such synchronized flows occur when the coupling strength is above a critical value. 
This critical coupling strength increases with oscillators' frequency detuning, i.e., the size ratio $r$ between the two cylinders. 
Thus, the region of synchronized flow in the $r$-$g$ map exhibits the shape of an inverse Arnold's tongue \citep{pikovsky2001universal}, as shown in figure \ref{fig:regimeMap}.
Note that the region of synchronization on the side of $r>1$ is slightly larger than that for $r<1$. 
This suggests that the wake coupling is stronger when the strength of vortex shedding is higher.

The bottom of the inverse Arnold's tongue is cut at $g \approx 1.2-1.4$, below which chaotic flows featured by irregular vortex shedding and unorganized lift phase trajectory are observed.
This is different from the scenario in some coupled oscillator systems where the Arnold's tongue expands infinitely with increasing coupling strength \citep{pikovsky2001universal}.
In the current case, the approaching of two side-by-side wakes results in a jet-like velocity profile in between.
The complex nonlinear interactions between these three instability mechanisms, i.e., two wakes and a jet, could lead to the chaotic flows.

\subsection{Spectral analysis}
\label{sec:spectral}

\begin{figure}
\centering
\includegraphics[width=0.95\textwidth]{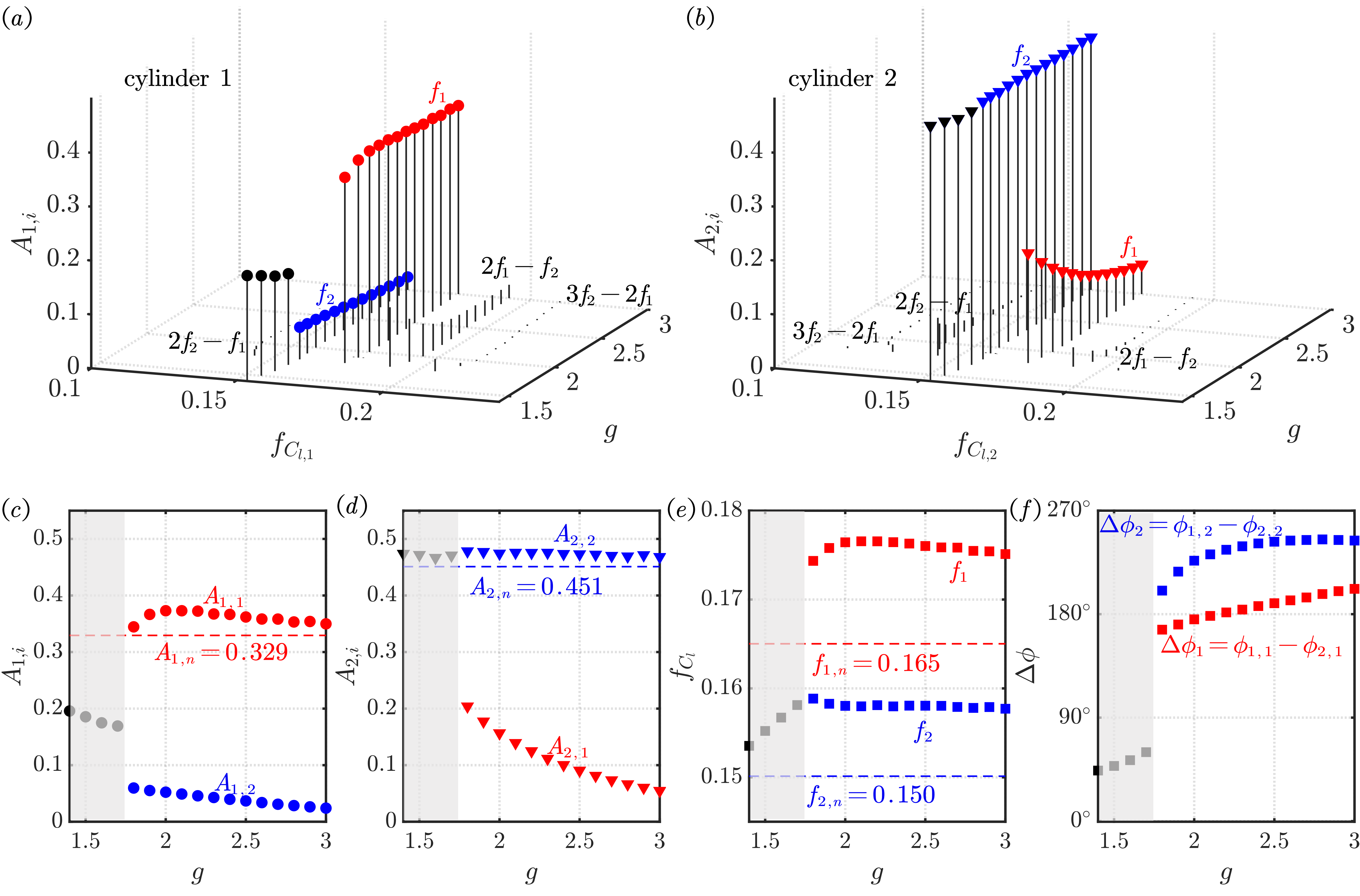}
\caption{Spectral analysis of lift coefficients of ($a$) cylinder 1 and ($b$) cylinder 2 for cases with $r=1.15$ using the NAFF algorithm. $(c)$ and ($d$): side views of $(a)$ and ($b$) showing amplitudes of lift coefficients on cylinders 1 and 2, respectively. Under the dashed lines, $A_{i,n}$ denotes the amplitude of lift coefficient (i.e., $C_{l,max}$) in the single cylinder case. $(e)$ top view of $(a)$ and ($b$) showing primary frequencies of the lift coefficients. $f_{i,n}$ denotes the natural shedding frequency in the wake of a single cylinder. $(f)$ phase difference of primary frequencies $f_1$ and $f_2$ between the two cylinders. In $(c)$-$(f)$, the shaded area represents the lock-in region. }
\label{fig:NAFF_r115}
\end{figure}

In this section, we focus on the quasi-periodic and synchronized flows to reveal how one wake affects its peer in a quantitative manner.
This is achieved through an advanced spectral analysis using the Numerical Analysis of Fundamental Frequencies (NAFF) algorithm \citep{laskar1990chaotic,laskar1999introduction}, which provides an accurate quasi-periodic approximation of the signal.
The algorithm starts by applying a Hanning window to the signal. 
Next, conventional Fast Fourier Transform (FFT) is applied to the signal to find the peaked frequency, which is used as the initial estimate for a numerical optimization of the overlap between the signal and $e^{-i2\pi ft}$ ($f$ is the frequency to be determined).
This allows determining $f$ as well as the corresponding amplitude and phase with high accuracy.
Once $f$ is determined, the overlap is subtracted from the original signal and the process is repeated to find the other frequencies.
The use of the NAFF algorithm avoids the spectral leakage problem that plagues the conventional FFT for quasi-periodic signals, and is critical for accurately quantifying the mutual interactions between the two wakes.

The amplitude spectra of the lift coefficients for cases with $r=1.15$ in the synchronized and quasi-periodic regimes are presented in figure \ref{fig:NAFF_r115}($a,b$).
For both cylinders, the lift coefficients are mainly composed of two primary frequencies denoted as $f_1$ and $f_2$, which are associated with the wakes of cylinders 1 and 2, respectively.
Thus, the two lift coefficients can be expressed as 
\begin{equation}
\begin{split}
  C_{l,1} &= A_{1,1}\sin(2\pi f_1 t + \phi_{1,1}) + A_{1,2}\sin(2\pi f_2 t + \phi_{1,2}) + ...,\\
	C_{l,2} &= A_{2,1}\sin(2\pi f_1 t + \phi_{2,1}) + A_{2,2}\sin(2\pi f_2 t + \phi_{2,2}) + ...,
\end{split}
\label{equ:NS}
\end{equation}
where $A_{i,j}$ is the amplitude associated with the frequency component $f_j$ on cylinder $i$, and $\phi_{i,j}$ is the corresponding phase. 
Here, $A_{1,1}$ and $A_{2,2}$ represent the strengths of vortex shedding of the two cylinders, and $A_{1,2}$ and $A_{2,1}$ characterize the mutual interference effects of the parallel wakes.
The phase differences associated with the two primary frequencies are defined as $\Delta \phi_1 = \phi_{1,1}-\phi_{2,1}$ and $\Delta \phi_2 = \phi_{1,2}-\phi_{2,2}$.

Over the quasi-periodic flows studied herein,  the strengths of vortex shedding for both cylinders ($A_{1,1}, A_{2,2}$) exhibit small variations with respect to the gap ratio, and tend towards their respective values in the single cylinder cases.
On the other hand, the primary frequencies $f_1$ and $f_2$ in the coupled wake are significantly larger than the natural shedding frequencies for both cylinders.
The mutual interference terms $A_{1,2}$ and $A_{2,1}$ are considerably smaller than the strengths of both wakes.
Both quantities decay as the gap ratio increases, indicating the weakening of mutual interactions.
It is also noted that due to nonlinear triad interactions between the two wakes, other frequency components with $mf_1 \pm nf_2$ (where $m$ and $n$ are integer numbers) known as sidebands \citep{craik1988wave,li2013lock} are also observed in the lift spectra for quasi-periodic flows, although their amplitudes are generally smaller than those of the primary frequencies.

The flow transitions from quasi-periodicity to synchronization as the gap ratio decreases below  1.8.
This is accompanied by abrupt drops in the strength and frequency of wake 1.
On the other hand, the strength and frequency for wake 2 exhibit small variations over the transition.
These observations suggest that the the process of synchronization is dominated by the stronger wake, to which the weaker wake is attracted.
Such master-slave coupling is ubiquitous in nonlinear interactions of nonidentical oscillators \citep{pikovsky2001universal}.
The transition to synchronization is also associated with abrupt jumps in the phase differences of the primary frequencies on both cylinders, as observed in figure \ref{fig:NAFF_r115}($f$).

\begin{figure}
    \centering
    \includegraphics[width=0.95\textwidth]{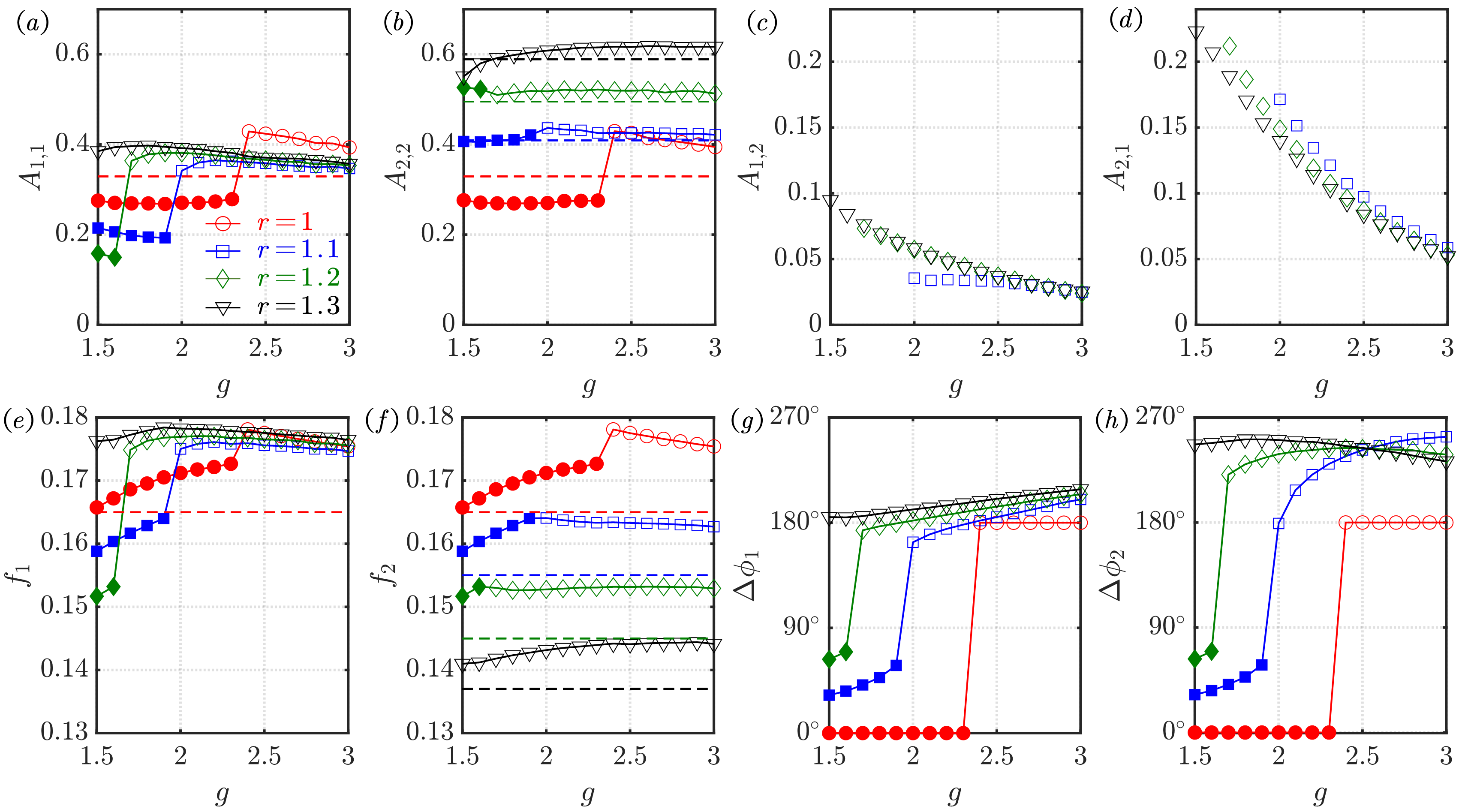}
    \caption{Key parameters in the spectra of lift coefficients for $r=1-1.3$ and $g=1.5-3$. Filled and empty symbols represent synchronized and quasi-periodic cases, respectively. The dashed lines indicate the quantities for the single cylinder cases.}
    \label{fig:NAFF_All}
\end{figure}

We further assess the effects of size ratio on  lift spectra in figure \ref{fig:NAFF_All}.
As expected, with growing diameter of cylinder 2, its strength of vortex shedding increases, and the frequency decreases.
In the quasi-periodic flows, the strength and frequency of vortex shedding for cylinder 1 both increase slightly increasing $r$.
The effect of cylinder 1 on wake 2 (indicated by $A_{2,1}$) weakens with increasing size ratio.
In contrast, the effect of cylinder 2 on wake 1 ($A_{1,2}$) is lower at $r=1.1$ than $r=1.2$ and 1.3, especially for cases with small gap ratios.
During synchronization, the frequency of wake 1 submits to that of wake 2, and the strength of wake 1 decreases with the size ratio considerably, resulting in large difference in the lift amplitudes between the ``master'' and ``slave'' oscillators.
The phase differences $\Delta \phi_1$ and $\Delta \phi_2$ are positively related with the size ratio $r$ during synchronization.
For quasi-periodic flows, the phase difference associated with $f_2$ is considerably higher than that with $f_1$.
Although omitted here, the above observations also apply to cases with $r<1$, for which the synchronization becomes dominated by cylinder 1.
From the above results, the in-phase and anti-phase synchronized flows observed in the $r=1$ cases can be regarded as special cases of the synchronized flow and the quasi-periodic flow, respectively.

\subsection{Modes of interaction}
\label{sec:DMD}

\begin{figure}
	\centering
	\includegraphics[width=0.95\textwidth]{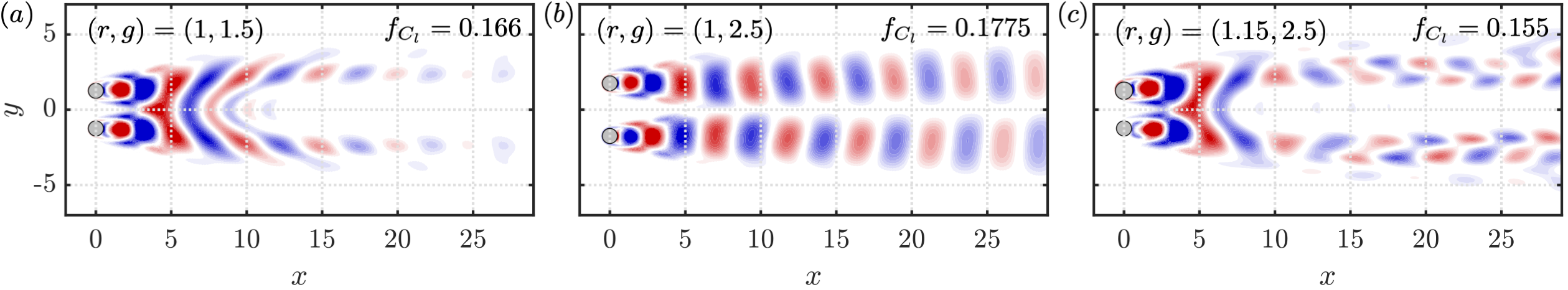}
	\caption{Dynamic modes for the synchronized wakes. $(a)$ $(r,g)=(1,1.5)$; $(b)$ $(r,g)=(1,2.5)$ and $(c)$ $(r,g)=(1.15,1.5)$. Blue color indicates negative vorticity, and red color indicate positive vorticity.}
	\label{fig:DMDSync}
\end{figure}

\begin{figure}
	\centering
	\includegraphics[width=0.99\textwidth]{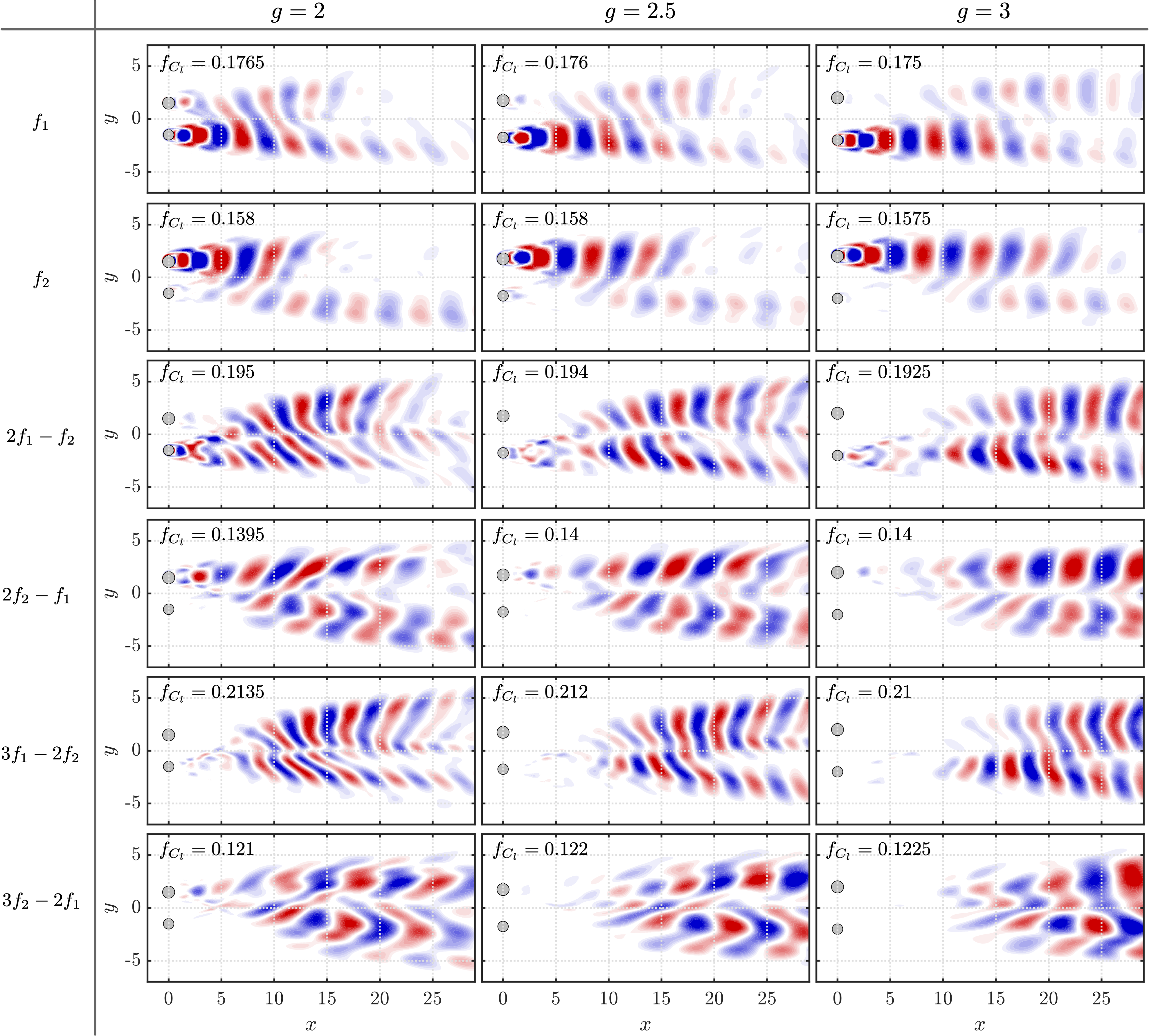}
	\caption{Dynamic modes for the quasi-periodic flows with $r=1.15$.}
	\label{fig:DMD115}
\end{figure}

We further perform dynamic mode decomposition \citep{schmid2010dynamic,rowley2009spectral} on the time-resolved vorticity fields to extract the coherent structures associated with the various frequency components in the coupled wakes. 
The dynamic modes for three types of synchronized wakes are shown in figure \ref{fig:DMDSync}.
For $(r,g)=(1,1.5)$, the in-phase synchronized flow features a pair of identical K\'arm\'an-like shedding mode in the near wake.
The two parallel wakes then merge at downstream, forming a single vortex street.
The anti-phase lock-in flow features two separate rows of K\'arm\'an wake structures, which are anti-symmetric with each other with respect to the $x$ axis.
For $r=1.15$, the synchronized wake mode at $g=1.5$ resembles the in-phase shedding mode of the case $(r,g)=(1,1.5)$. 
However, due to the size difference, the mode is not perfectly symmetric along the $x$ axis.

The dynamic modes for the quasi-periodic flows exhibit rich features as shown in figure \ref{fig:DMD115}.
The primary frequencies $f_1$ and $f_2$ are associated with typical K\'arm\'an modes in the near wake of each cylinder, with their imprints also appearing in the neighbouring wake.
The signature of the $f_1$ mode in the wake of cylinder 2 is stronger, and appears closer to the bodies, than that of the $f_2$ mode in cylinder 1's wake. 
This is in agreement with the observation in figure \ref{fig:NAFF_r115} that $A_{2,1}$ is larger than $A_{1,2}$.
With growing gap ratio, these imprints shift downstream, suggesting the weakening of mutual interactions.
These primary modes are strong in the near wake and decay towards far wake.

The dynamic modes associated with the sideband frequencies $(mf_1+nf_2)$ feature coherent structures of varying spatial scales. 
The lower frequency corresponds to larger-scale structures, and high frequency the finer.
Unlike the primary modes with frequencies $f_1$ and $f_2$ that are prevalent in the near wake, the dynamic modes of the sideband frequencies are weak in near wake, and grows downstream to dominate the far wake, especially for cases with large gap ratios.
As a result, these sideband frequencies are less felt in the lift forces, as observed in figure \ref{fig:NAFF_r115}($a,b$).
The nonlinear interaction between the two nonidentical cylinder wakes breeds a cascade of dynamic modes with various spatio-temporal scales as the flow convects downstream, and increases the complexity of the far wake in quasi-periodic flows.

\section{Conclusion}
\label{sec:conclusion}
We have systematically studied flows over a pair of side-by-side circular cylinders with different sizes at a fixed Reynolds number of 100 using direct numerical simulations.
The aim is to understand how the parallel wakes affect each other in a quantitative manner.
Over a range of size ratios and gap ratios, the coupled wakes exhibit distinct dynamical features including the quasi-periodic flows for cases with large gap ratio, chaotic flow for cases with small gap ratio, and synchronized flows in between taking the shape of an inverse Arnold's tongue in the $r$--$g$ space.
To quantify the mutual interactions of the parallel wakes, spectral analysis and dynamic mode decomposition are performed to extract the frequencies and the associated amplitudes, phases, and spatial modes from the flows.
We show that the quasi-periodic flows are mainly featured by two primary frequencies generated from the vortex streets behind the two cylinders.
The nonlinear interactions between two vortex streets not only leave signature of each primary mode in the wake of the neighbouring cylinder, but also breed the triad sideband frequencies that become more active in the far wake.
Over the transition from the quasi-periodicity to synchronization, the vortex shedding behind the smaller cylinder submits to the larger's wake, while the latter remains almost unchanged in terms of shedding frequency and strength.
The new insights obtained from this study improve the current understanding of flows over two side-by-side cylinders, and are important for interpreting more complicated wake interactions problems.

\vspace{0.2cm}

{\small \textbf{Acknowledgement}.
KZ is grateful for the Office of Advanced Research Computing (OARC) at Rutgers University for providing access to the Amarel cluster. 

\vspace{0.2cm}

\textbf{Declaration of interest}.
The authors report no conflict of interest.
}
\bibliography{reference}

\begin{thebibliography}{25}
\expandafter\ifx\csname natexlab\endcsname\relax\def\natexlab#1{#1}\fi
\def\au#1{#1} \def\ed#1{#1} \def\yr#1{#1}\def\at#1{#1}\def\jt#1{\textit{#1}}
  \def\bt#1{#1}\def\bvol#1{\textbf{#1}} \def\vol#1{#1} \def\pg#1{#1}
  \def\publ#1{#1}\def\arxiv#1{#1}\def\org#1{#1}\def\st#1{\textit{#1}}

\bibitem[Bearman \& Wadcock(1973)]{bearman1973interaction}
{\sc \au{Bearman, P.~W.} \& \au{Wadcock, A.~J.}} \yr{1973}  \at{The interaction
  between a pair of circular cylinders normal to a stream}.  \jt{J. Fluid
  Mech.}  \bvol{61}~(3),  \pg{499--511}.

\bibitem[Carini {\em et~al.\/}(2014)Carini, Giannetti \&
  Auteri]{carini2014origin}
{\sc \au{Carini, M.}, \au{Giannetti, F.} \& \au{Auteri, F.}} \yr{2014}  \at{On
  the origin of the flip--flop instability of two side-by-side cylinder wakes}.
   \jt{J. Fluid Mech.}  \bvol{742},  \pg{552--576}.

\bibitem[Craik(1988)]{craik1988wave}
{\sc \au{Craik, A. D.~D.}} \yr{1988} {\em Wave interactions and fluid flows\/}.
   \publ{Cambridge University Press}.

\bibitem[Henderson(1995)]{henderson1995details}
{\sc \au{Henderson, R.~D.}} \yr{1995}  \at{Details of the drag curve near the
  onset of vortex shedding}.  \jt{Phys. Fluids}  \bvol{7}~(9),
  \pg{2102--2104}.

\bibitem[Inoue \& Suzuki(2008)]{inoue2008beat}
{\sc \au{Inoue, O.} \& \au{Suzuki, Y.}} \yr{2008}  \at{Beat of sound generated
  by flow past two side-by-side square cylinders of different sizes}.
  \jt{Phys. Fluids}  \bvol{20}~(7),  \pg{076101}.

\bibitem[Kang(2003)]{kang2003characteristics}
{\sc \au{Kang, S.}} \yr{2003}  \at{Characteristics of flow over two circular
  cylinders in a side-by-side arrangement at low {R}eynolds numbers}.
  \jt{Phys. Fluids}  \bvol{15}~(9),  \pg{2486--2498}.

\bibitem[Kim \& Durbin(1988)]{kim1988investigation}
{\sc \au{Kim, H.-J.} \& \au{Durbin, P.~A.}} \yr{1988}  \at{Investigation of the
  flow between a pair of circular cylinders in the flopping regime}.  \jt{J.
  Fluid Mech.}  \bvol{196},  \pg{431--448}.

\bibitem[Laskar(1990)]{laskar1990chaotic}
{\sc \au{Laskar, J.}} \yr{1990}  \at{The chaotic motion of the solar system: A
  numerical estimate of the size of the chaotic zones}.  \jt{Icarus}
  \bvol{88}~(2),  \pg{266--291}.

\bibitem[Laskar(1999)]{laskar1999introduction}
{\sc \au{Laskar, J.}} \yr{1999}  \at{Introduction to frequency map analysis}.
  \bt{In {\em Hamiltonian systems with three or more degrees of freedom\/}},
  \pg{pp. 134--150}.  \publ{Springer}.

\bibitem[Lee {\em et~al.\/}(2009)Lee, Yang \& Y.]{lee2009flow}
{\sc \au{Lee, K.-J.}, \au{Yang, K.-S.} \& \au{Y., D.-H.}} \yr{2009}
  \at{Flow-induced forces on two circular cylinders in proximity}.  \jt{Comput.
  Fluids}  \bvol{38}~(1),  \pg{111--120}.

\bibitem[Li \& Juniper(2013)]{li2013lock}
{\sc \au{Li, L. K.~B.} \& \au{Juniper, M.~P.}} \yr{2013}  \at{Lock-in and
  quasiperiodicity in a forced hydrodynamically self-excited jet}.  \jt{J.
  Fluid Mech.}  \bvol{726},  \pg{624--655}.

\bibitem[Mizushima \& Ino(2008)]{mizushima2008stability}
{\sc \au{Mizushima, J.} \& \au{Ino, Y.}} \yr{2008}  \at{Stability of flows past
  a pair of circular cylinders in a side-by-side arrangement}.  \jt{J. Fluid
  Mech.}  \bvol{595},  \pg{491--507}.

\bibitem[Norberg(2003)]{norberg2003fluctuating}
{\sc \au{Norberg, C.}} \yr{2003}  \at{Fluctuating lift on a circular cylinder:
  review and new measurements}.  \jt{J. Fluids Struct.}  \bvol{17}~(1),
  \pg{57--96}.

\bibitem[Octavianty \& Asai(2015)]{octavianty2015synchronized}
{\sc \au{Octavianty, R.} \& \au{Asai, M.}} \yr{2015}  \at{Synchronized vortex
  shedding and sound radiation from two side-by-side rectangular cylinders of
  different cross-sectional aspect ratios}.  \jt{Phys. Fluids}  \bvol{27}~(10),
   \pg{107103}.

\bibitem[Peschard \& Le~Gal(1996)]{peschard1996coupled}
{\sc \au{Peschard, I.} \& \au{Le~Gal, P.}} \yr{1996}  \at{Coupled wakes of
  cylinders}.  \jt{Phys. Rev. Lett.}  \bvol{77}~(15),  \pg{3122}.

\bibitem[Pikovsky {\em et~al.\/}(2001)Pikovsky, Rosenblum \&
  Kurths]{pikovsky2001universal}
{\sc \au{Pikovsky, A.}, \au{Rosenblum, M.} \& \au{Kurths, J.}} \yr{2001} {\em
  Synchronization: A Universal Concept in Nonlinear Sciences\/}.
  \publ{Cambridge University Press}.

\bibitem[Ren {\em et~al.\/}(2021)Ren, Cheng, Xiong, Tong \&
  Chen]{ren2021bistabilities}
{\sc \au{Ren, C.}, \au{Cheng, L.}, \au{Xiong, C.}, \au{Tong, F.} \& \au{Chen,
  T.}} \yr{2021}  \at{Bistabilities in two parallel {K}{\'a}rm{\'a}n wakes}.
  \jt{J. Fluid Mech.}  \bvol{929}.

\bibitem[Rowley {\em et~al.\/}(2009)Rowley, Mezi{\'c}, Bagheri, Schlatter \&
  Henningson]{rowley2009spectral}
{\sc \au{Rowley, C.~W.}, \au{Mezi{\'c}, I.}, \au{Bagheri, S.}, \au{Schlatter,
  P.} \& \au{Henningson, D.~S.}} \yr{2009}  \at{Spectral analysis of nonlinear
  flows}.  \jt{J. Fluid Mech.}  \bvol{641},  \pg{115--127}.

\bibitem[Schmid(2010)]{schmid2010dynamic}
{\sc \au{Schmid, P.~J.}} \yr{2010}  \at{Dynamic mode decomposition of numerical
  and experimental data}.  \jt{J. Fluid Mech.}  \bvol{656},  \pg{5--28}.

\bibitem[Sumner(2010)]{sumner2010two}
{\sc \au{Sumner, D.}} \yr{2010}  \at{Two circular cylinders in cross-flow: {A}
  review}.  \jt{J. Fluids Struct.}  \bvol{26}~(6),  \pg{849--899}.

\bibitem[Weller {\em et~al.\/}(1998)Weller, Tabor, Jasak \&
  Fureby]{weller1998tensorial}
{\sc \au{Weller, H.~G.}, \au{Tabor, G.}, \au{Jasak, H.} \& \au{Fureby, C.}}
  \yr{1998}  \at{A tensorial approach to computational continuum mechanics
  using object-oriented techniques}.  \jt{Comput. Phys.}  \bvol{12}~(6),
  \pg{620--631}.

\bibitem[Williamson(1985)]{williamson1985evolution}
{\sc \au{Williamson, C. H.~K.}} \yr{1985}  \at{Evolution of a single wake
  behind a pair of bluff bodies}.  \jt{J. Fluid Mech.}  \bvol{159},
  \pg{1--18}.

\bibitem[Williamson(1996)]{williamson1996vortex}
{\sc \au{Williamson, C. H.~K.}} \yr{1996}  \at{Vortex dynamics in the cylinder
  wake}.  \jt{Annu. Rev. Fluid Mech.}  \bvol{28}~(1),  \pg{477--539}.

\bibitem[Zdravkovich(1997)]{zdravkovich1997flow}
{\sc \au{Zdravkovich, M.~M.}} \yr{1997} {\em Flow around circular cylinders,
  volume 2: Applications\/}.  \publ{Oxford university press}.

\bibitem[Zhou \& Alam(2016)]{zhou2016wake}
{\sc \au{Zhou, Y.} \& \au{Alam, M.~M.}} \yr{2016}  \at{Wake of two interacting
  circular cylinders: {A} review}.  \jt{Int. J. Heat Fluid Flow}  \bvol{62},
  \pg{510--537}.

\end{thebibliography}
\bibliographystyle{jfm}
\end{document}